\documentclass[preprints,communication,accept,moreauthors,pdftex]{Definitions/mdpi}

\firstpage{1} 
\pubvolume{5}
\issuenum{5}
\articlenumber{0}
\pubyear{2019}
\copyrightyear{2019}
\history{}
\usepackage{booktabs} 
\usepackage{multirow}
\usepackage{soul} 
\usepackage{microtype}
\setitemize{parsep=6pt,itemsep=0pt,leftmargin=*,labelsep=5.5mm}
\setenumerate{parsep=6pt,itemsep=0pt,leftmargin=*,labelsep=5.5mm}
\setstcolor{red}

\Title{\textls[-20]{Azimuthal correlations of D mesons with charged particles in simulations with the ALICE experiment}}

\Author{Eszter Frajna $^{1,2,}$*\orcidA{} and Robert Vertesi $^{1}\orcidB{}$ for the ALICE Collaboration}

\AuthorNames{Eszter Frajna, Robert Vertesi [ALICE]}

\address{
$^{1}$ \quad Wigner Research Centre for Physics, Konkoly-Thege Mikl\'{o}s \'{u}t 29-33, 1121 Budapest, Hungary\\
$^{2}$ \quad Department of Physics, Budapest University of Technology and Economics, M\H uegyetem rkp. 3, 1111~Budapest,~Hungary}

\corres{Correspondence: frajna.eszter@wigner.hu}
\firstnote{This paper is based on the talk at the ACHT 2021 "Perspectives in Particle,  Cosmo- and Astroparticle Theory", online, 21-23 April 2021.} 

\abstract {In this manuscript, results of a component-level analysis with Monte Carlo simulations are presented, that aid the interpretation of recent ALICE results on the azimuthal correlation distribution of prompt D mesons with charged hadrons in pp and p--Pb collisions at $\sqrt{s_{\rm NN}}$ = 5.02 TeV. Parton-level contributions and fragmentation properties are evaluated. Charm and beauty contributions are compared in order to identify the observables that serve as sensitive probes of the production and hadronization of heavy quarks.}


\begin{document}

\section{Introduction}

Investigating the characteristic correlation patterns of heavy-flavour (charm and beauty) particles in ultra-relativistic hadronic collisions can help to understand the flavour-dependence of the fragmentation mechanisms. 
The angular-correlation between prompt D mesons and charged particles in proton--proton (pp) collisions is sensitive to the perturbative quantum-chromodynamics (QCD) mechanisms of charm-quark production, as well as the fragmentation into charm hadrons, and are also affected by initial and final-state parton radiations (ISR and FSR) as well as by multi-parton interactions (MPI).

With the ALICE detector, the heavy-flavour hadrons are reconstructed from their decay products. In order to investigate the mechanisms of particle production and study the event properties of ultra-relativistic hadronic collisions in ALICE, two-particle angular correlations are used. They are especially effective in the $p_{\mathrm{T}}$ ranges ($p_{\rm T} \sim $ 1--3 GeV/$c$) where jet-medium interactions and semi-hard QCD processes are present, and a full jet reconstruction may be problematic \cite{Adam:2016tsv}. Heavy-flavour correlation measurements in pp collisions serve as a reference for possible flavour-dependent modification of production and fragmentation by hot or cold nuclear matter in larger systems \cite{Jung:2014qta}.

ALICE reported the azimuthal correlation distributions of prompt D mesons with charged particles in pp and p–Pb collisions at $\sqrt{s_{\rm NN}}=5.02$ TeV \cite{D-h}. The transverse momentum  ($p_{\mathrm{T}}$) evolution of the near- and away-side peaks are found to be consistent in pp and p–Pb collisions, in the whole $p_{\mathrm{T}}$ range. This suggests that the fragmentation and hadronisation of charm quarks is not strongly influenced by cold-nuclear-matter effects. The baseline-subtracted correlation functions and the near- and away-side peak yields and widths measured in pp collisions are compared to predictions by several event generators, with different modelling of charm production, parton showering, and hadronisation. In general, the models describe the main features of the correlation functions well.

In the current manuscript we present a component-level analysis based on simulations performed within the ALICE software framework matched to the ALICE kinematic range, following the analysis procedure of Ref. \cite{D-h}. In order to deepen the physical understanding on the development of correlation distributions via the production, parton shower and fragmentation in heavy-flavour events, the effect of FSR and ISR as well as the MPI are investigated. We also explore the effects of quark mass and the different fragmentation models. Correlations of prompt and non-prompt D mesons to charged hadrons are also compared in order to quantify expected differences between different flavours.

\section{Analysis Method}
The $p_{\rm T}$ dependence  of  the  correlation  is  studied  by  measuring  triggered  correlations. The D-mesons are defined as trigger particles, while associated particles are considered as all charged primary particles (see Ref. \cite{D-h} for a more complete description of the analysis procedure). Non strange D mesons were reconstructed in the central rapidity region from their charged hadronic decay channels ${\rm D}^0 \rightarrow {\rm K}^-\pi^+$, ${\rm D}^+\rightarrow {\rm K}^-\pi^+\pi^+$, ${\rm D}^{*+} \rightarrow {\rm D}^0 \pi^+ $, and charge conjugates.

The associated particles are correlated to the trigger particle. The associated per-trigger yield is measured as a function of the azimuthal angle difference of the D meson ($\varphi_{\rm D}$) and the associated particle ($\varphi_{\rm assoc.}$) as $\Delta \varphi = \varphi_{\rm D}-\varphi_{\rm assoc.}$, in several ranges of the trigger transverse momentum ($p_{\rm T}^{\rm D}$) and associated transverse momentum ($p_{\rm T}^{\rm assoc.}$) within each event, where $p_{\rm T}^{\rm assoc.}< p_{\rm T}^{\rm D}$. The near-side peak is defined as the range from $-\frac{\pi}{2}$ to $\frac{\pi}{2}$, and the away-side as the range from $\frac{\pi}{2}$ to $\frac{3\pi}{2}$. Near-side correlations give information about the structure of jets since, after background subtraction, most trigger and associated particles come from the same jet. Differences between correlations of beauty, charm and light flavour provide valuable information about flavour-dependent jet fragmentation. Away-side correlation is mostly from back-to-back jet pairs and is sensitive to the underlying hard processes.

We used the PYTHIA8 Monte Carlo event generator \cite{Sjostrand:2007gs} to simulate hard QCD events using the 4C \cite{Corke:2010yf} tune for LHC pp data at $\sqrt{s}=5.02$ TeV. In general, PYTHIA8 with 4C tune describe the trends in ALICE D--h correlation data well with some qualitative differences \cite{D-h}.

Charged hadrons are selected in $|\eta|<0.8$, with a similar method to that outlined in Ref.~\cite{D-h}. The azimuthal-correlation distributions are examined in the pseudorapidity-difference range $|\Delta\eta|<$ 1 for four D-meson $p_{\rm T}$ ranges, $3 < p_{\rm T}^{\rm D}< 5$ GeV/$c$, $5 < p_{\rm T}^{\rm D} < 8$ GeV/$c$, $8 < p_{\rm T}^{\rm D}< 16$ GeV/$c$, and $16 < p_{\rm T}^{\rm D} < 24$ GeV/$c$. The distributions are presented for $p_{\rm T}^{\rm assoc.} > 0.3$ GeV/$c$, as well as for three sub-ranges, $0.3 < p_{\rm T}^{\rm assoc.} < 1$ GeV/$c$, $1 < p_{\rm T}^{\rm assoc.}< 2$ GeV/$c$, and $2 < p_{\rm T}^{\rm assoc.} < 3$ GeV/$c$.

In order to quantify the properties of the average D-meson azimuthal-correlation distribution, the latter was fitted with the following function:

\begin{equation} \label{eq:fit}
f(\Delta \varphi)= b+ \frac{Y_{NS} \cdot \beta}{2\alpha\Gamma(1/\beta)} \cdot e^{-(\frac{\Delta \varphi}{\alpha})^{\beta}}+ \frac{Y_{AS}}{\sqrt{2\pi}\sigma_{AS}} \cdot e^{-\frac{(\Delta \varphi-\pi)^2}{2\sigma_{AS}^2}} \ .
\end{equation}

The fit function in Eq. \ref{eq:fit} is composed of a constant term b describing the flat contribution below the correlation peaks, a generalised Gaussian term describing the near-side peak, and a Gaussian reproducing the away-side peak. In the generalised Gaussian, the term $Y_{\mathrm{NS}}$ is a normalization factor for the near-side peak, the term $\alpha$ is related to the variance of the function, hence to its width, while the term $\beta$ drives the shape of the peak (if  $\beta$ = 1, then it is an exponential function, the Gaussian function is obtained from $\beta$ = 2 and if $\beta>2$, the shape of the peak is flattened). In the Gaussian, the term $Y_{\mathrm{AS}}$ and $\sigma_{\mathrm{AS}}$ are the yield and the width, respectively, of the away-side peak. 


\section{Results} \label{sec:Results}

\subsection{Different parton level contributions}

We examined the three different parton-level contributions implemented in PYTHIA 8: MPI, ISR and FSR \cite{MPI, ISRandFSR}. In PYTHIA8, multiple hard-parton interactions and scatterings between proton remnants are modelled in the multi-parton interaction framework. These are responsible for the production of a large fraction of the particles uncorrelated with the D-meson candidate trigger. Before hard scattering takes place, one of the incoming partons radiates gluons in the so-called initial-state radiation process. Similarly, outgoing partons from the hard-scattering produces a shower of softer particles via a final-state radiation process.

\begin{figure}[H]
    \centering
	\includegraphics[width=11cm]{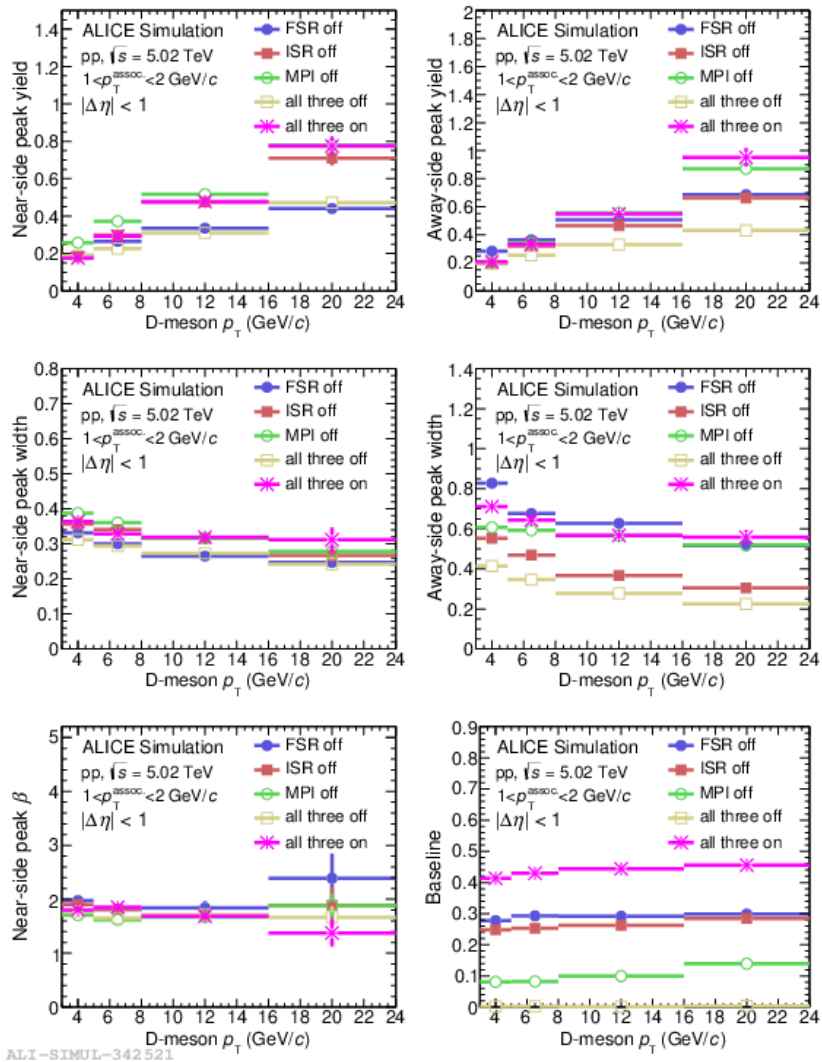}
	\caption{Near- and away-side peak yield (top row), width (middle row), $\beta$ parameter and baseline (bottom row) of D--h correlations from simulations with different parton level contributions in pp collisions at $\sqrt{s}= 5.02$ TeV, as a function of the D-meson $p_{\rm T}$, for  $ 1<p_{\rm T}^{\rm assoc.}<2$ GeV/$c$.}
	\label{fig:parton}
\end{figure}

\begin{figure}[H]
	\centering
	\includegraphics[width=9cm]{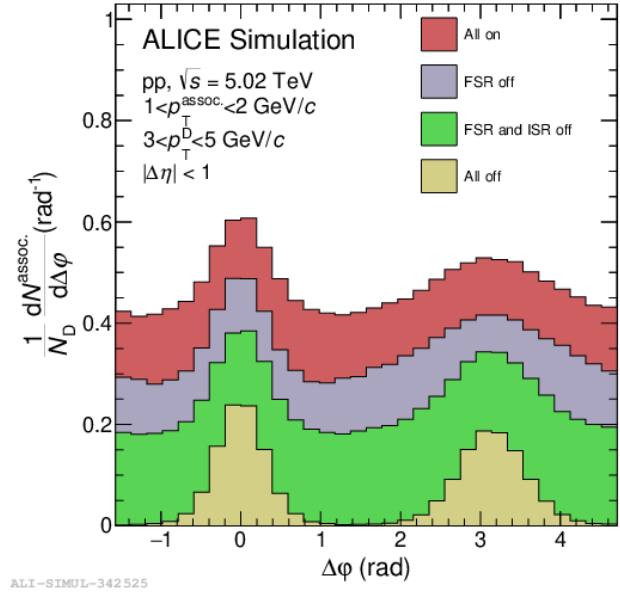}
	\caption{Associated yield of D--h correlations from simulations with different parton level contributions of prompt D mesons with hadrons in simulated pp collisions at $\sqrt{s}= 5.02$ TeV, for $ 1<p_{\rm T}^{\rm assoc.}<2$ GeV/$c$ and $3< p_{\rm T}^{\rm D}< 5 $GeV/$c$.}
    \label{fig:parton1}
\end{figure}
 
By switching off/on the different interactions at the parton level while simulating PYTHIA 8 events, it is possible to study the effect of these interactions on the observed correlations. Figure \ref{fig:parton} compares the near-side peak features and $p_{\rm T}$ evolution of data with simulated events with the presence of different interactions for $1<p_{\rm T}^{\rm assoc.}<2$ GeV/$c$, as a function of $p_{\rm T}^{\rm D}$. In the top row, the near-side yield shows significant contribution of FSR at higher $p_{\rm T}^{\rm D}$, while the away-side width demonstrates significant contribution from both FSR and ISR. In the middle row, the near-side width and shape display no change, which attests the shape of the near-side peak is driven by the fragmentation and the hadronic stage. As expected, the away-side peak is wider than the near-side peak because of a combined contribution of parton-level effects (with ISR having the strongest role). When FSR is turned off, it overshoots the physical case (all on). Figure \ref{fig:parton1} shows an example of the associated yield of D--h correlations from simulations with different parton level contributions. The observation that ISR does not influence the near-side peak observables, can probably be understood considering that these early parton radiations tend not to influence jet development with a given $p_{\rm T}$ but they may statistically shift the $p_{\rm T}$ interval in which the paired jet falls. In the bottom row the baseline presents the contributions of parton-level effects to the underlying event as expected. 

Although all three effects contribute to the baseline, the most pronounced one is MPI, and the observed $p_{\rm T}^{\rm D}$ dependence is not significant. This is expected since MPI in high-multiplicity pp collisions is proposed to explain the observed trend of D-meson self-normalised yields at different relative multiplicities, which is largely independent from the leading hard process. 

\subsection{Investigating the fragmentation of heavy-flavour}

Identifying characteristic correlation patterns of heavy quarks can help understand flavour-dependent fragmentation. Measuring the yield of heavy quarks provides insight into the perturbatively computable processes of flavour-dependent parton formation, while the study of the jet structures also provides information on fragmentation processes that cannot be calculated analytically. The azimuthal correlation of D mesons with light charged hadrons reveals information of the formation of the jet structure. In the following, two heavy-flavour fragmentation models, the Lund and the Peterson models \cite{Andersson:1986au, Peterson} are compared. The Lund model is a phenomenological model of hadronization that has been used in the field since its emergence. It treats all but the highest-energy gluons as field lines, which are attracted to each other due to the gluon self-interaction and so form a narrow tube of strong colour field. The Lund formula is
\begin{equation}\label{eq:peterson}
f(z) = \frac{(1-z)^a}{z} \mathrm{exp} \bigg(-\frac{bm^2_{\perp h}}{z}\bigg)~,
\end{equation}
where $f(z)$ is the parton density in colliding hadron, $z$ is the momentum fraction, $m_\perp= \sqrt{m^2+p_\perp^2}$ is the transverse mass and $a$ and $b$ are parameters. In accordance with observations, the model predicts that in addition to the particle jets formed along the original paths of two separating quarks, there will be a spray of hadrons produced between the jets. The Peterson formula,
\begin{equation}\label{eq:peterson}
f(z) = \frac{1}{z(1-\frac{1}{z}-\frac{\epsilon}{1-z})^2}~, 
\end{equation}
where $\epsilon$ is the ratio of the effective light and heavy quark masses, is a simple parametrization of the heavy-quark fragmentation functions that describes well the charmed hadron spectra measured at SLAC. 

\begin{figure}[H]
	\centering
	\begin{minipage}[t]{.49\textwidth}
		\centering
		\includegraphics[width=6.7cm,height=6cm]{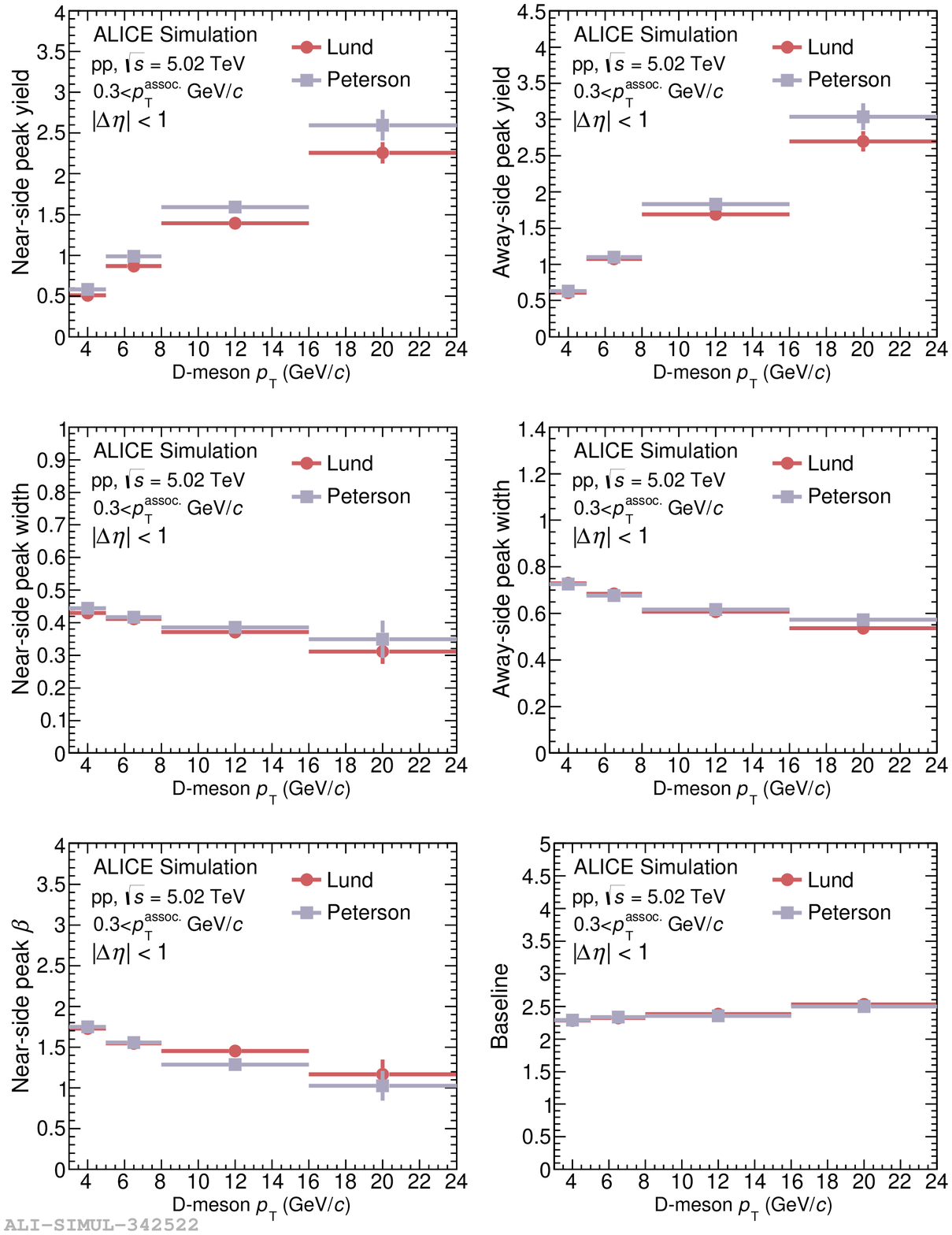}
		\caption{Near-side peak yield of simulated D--h correlations with different heavy-flavour fragmentation models in pp collision at $\sqrt{s}= 5.02$ TeV, as a function of the D-meson $p_{\rm T}$, for  $ p_{\rm T}^{\rm assoc.}> 0.3$ GeV/$c$.}
		\label{fig:peterson}
	\end{minipage}
	\hfill
	\begin{minipage}[t]{.49\textwidth}
		\centering
		\includegraphics[width=6.7cm,height=6cm]{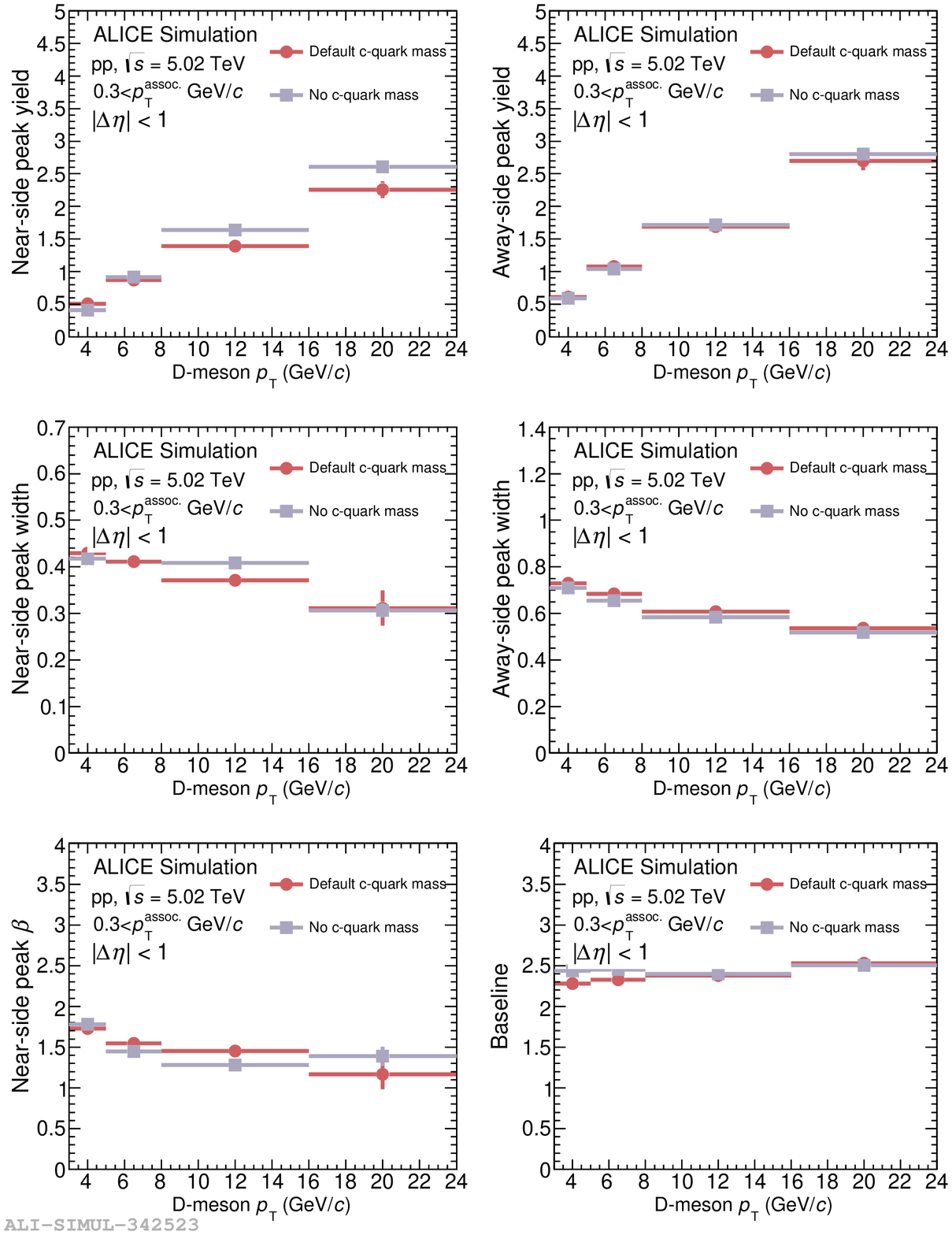}
		\caption{Near-side peak yield of simulated D--h correlations in pp collision at $\sqrt{s}= 5.02$ TeV, as a function of the D-meson $p_{\rm T}$, for  $ p_{\rm T}^{\rm assoc.}> 0.3$ GeV/$c$.}
        \label{fig:nomass}
	\end{minipage}%
\end{figure}

Figure \ref{fig:peterson} compares the near-side peak yields and the $p_{\rm T}$ evolution for the Lund and the Peterson heavy-flavour fragmentation models in pp collisions, by the PYTHIA 8 event generator, for $0.3 <p_{\rm T}^{\rm assoc.}$ GeV/$c$ and for different $p_{\rm T}^{\rm D}$ ranges. It presents hints of different trends towards higher D-meson $p_\mathrm{T}$.

The mass of the parton initiating the jet also influences the structure of the resulting hadronic jets. When charged particles are produced in high-energy collisions, they are usually accompanied by FSR. In the heavy quark case QCD predicts a suppression of soft gluon emission within a cone with a given angle around the moving direction of a charged particle. This is called the dead-cone effect \cite{deadcone1}. The pattern of radiation depends crucially on the mass of the emitter but not on its spin, where radiation from quarks with mass $m$ and energy $E$ is suppressed for emission angles $\theta \leq m/E$. The first direct measurement at the LHC of the dead-cone effect was carried out recently by ALICE \cite{deadcone3}. However, we also expect to see a modified fragmentation in the correlations, due to the dead-cone effect.

In order to disentangle the dead-cone effect from other differences that may show up between heavy- and light-flavour correlations, the fragmentation mass of the charm quark was set to zero. In Fig. \ref{fig:nomass} there are slight differences at the near-side yield and at high $p_{\rm T}^{\rm D}$ in a similar pattern to that in Fig.~\ref{fig:peterson}. This confirms that the differences in fragmentation of light and heavy quarks are accessible with angular correlation measurements.

\subsection{Separation of prompt and non-prompt D mesons}

\begin{figure}[H]
	\centering
	\includegraphics[width=11cm]{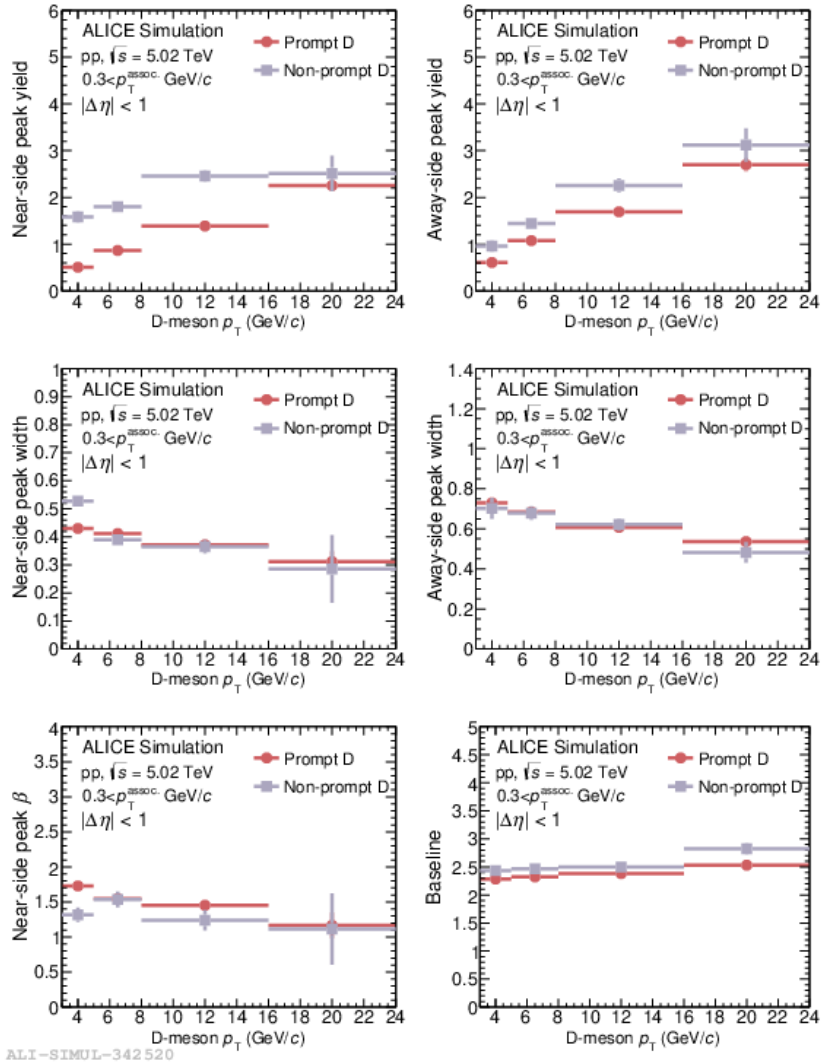}
	\caption{Near- and away-side peak yield (top row), width (middle row), $\beta$ parameter and baseline (bottom row) of prompt and non-prompt D meson correlations with hadrons in simulated pp collisions at $\sqrt{s}= 5.02$ TeV, as a function of the D-meson $p_{\rm T}$, for  $ p_{\rm T}^{\rm assoc.}> 0.3$ GeV/$c$. }
	\label{fig:prompts}
\end{figure}
\begin{figure}[H]
	\centering
	\includegraphics[width=9cm]{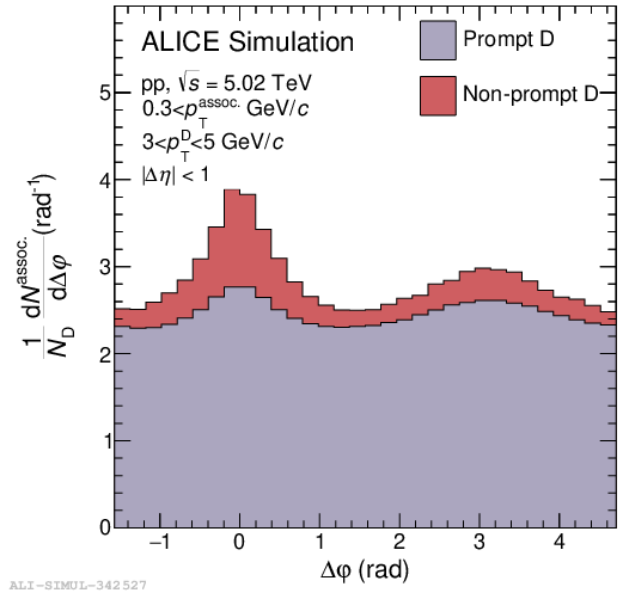}
	\caption{Associated yield of prompt and non-prompt D meson correlations with hadrons in simulated pp collisions at $\sqrt{s}= 5.02$ TeV, for  $ p_{\rm T}^{\rm assoc.}> 0.3$ GeV/$c$ and $3< p_{\rm T}^{\rm D}< 5 $GeV/$c$. }
    \label{fig:prompt}
\end{figure}

We examined the direct decay cases of D mesons  promptly produced from c quarks and non-prompt D mesons from the decay of B mesons. The measurement of the relative contributions of prompt and non-prompt D mesons allow for the verification of QCD models in pp collisions, as well as and for the understanding of the flavour-dependence of heavy-quark energy loss within the quark-gluon plasma in heavy-ion collisions. Figure \ref{fig:prompts} shows a strong difference between trends in the parameters of prompt and non-prompt D mesons and Fig. \ref{fig:prompt} shows an example of the associated yield of prompt and non-prompt D meson correlations. The per-trigger yields are significantly higher in case of non-prompt D mesons at both the near and away sides (about 50$\%$) and for the baseline (about 10$\%$), meaning a higher particle production associated to the heavier B mesons.

The near-side shapes are also significantly different at low $p_{\rm T}^{\rm D}$ similarly to the heavy-flavour electrons from b and c quarks \cite{Frajna}. This attests to the importance of flavour-dependent fragmentation in the resulting correlation patterns.

\section{Conclusions}

Recent ALICE results on the azimuthal correlation distribution of prompt D mesons with charged hadrons in pp and p--Pb collisions at $\sqrt{s_{\rm NN}}$ = 5.02 TeV are qualitatively described by several LO and NLO simulations \cite{D-h}. Among these, PYTHIA 8, which includes MPI and colour reconnection description, well describes the measured correlations. For this reason we performed a component level analysis with PYTHIA 8 to amend and interpret these measurements.

We observe that the contribution of parton-level effects mostly affect the underlying event and away-side peak. While the away-side width is sensitive to the MPI contribution, the baseline is affected by the ISR, FSR and MPI. The difference in the correlation functions depending on whether the Lund or the Peterson fragmentation model is used, comes from the effect of the charm mass. There are slight differences in the correlation patterns when setting the c-quark mass to zero, showing the role of dead-cone effect in fragmentation. The near-side peaks of prompt and non-prompt D mesons are significantly different, highlighting the importance of flavour-dependent fragmentation in the observed correlation patterns.

The results obtained from the analysis help identifying the observables that serve as sensitive probes of the production and hadronisation of heavy quarks produced in ultra-relativistic pp collisions. Moreover, they provide ground for the interpretation of possible modifications of the production and fragmentation in heavy-ion systems.

\vspace{6pt} 





\section*{Acknowledgements}

\funding{This work has been supported by the Hungarian National Research Fund (OTKA) grant FK131979, K135515 and 2019-2.1.11-TÉT-2019-00050.}

\conflictsofinterest{The author declares no conflict of interest.}



\reftitle{References}





\end{document}